 \documentclass[final,5p,times,twocolumn]{elsarticle}

 \usepackage{graphics}
 \usepackage{graphicx}

\usepackage{amssymb}
\usepackage{amsmath}

\journal{Journal of Molecular Spectroscopy}

\begin{document}

\begin{frontmatter}



\title{Spectroscopic parameters for silacyclopropynylidene, SiC$_2$, from extensive 
astronomical observations toward CW~Leo (IRC~+10216) with the {\it Herschel} 
satellite\tnoteref{Widmung}\tnoteref{Herschel-Acknowledgment}}
\tnotetext[Widmung]{We dedicate this work to the memory of Gisbert Winnewisser, 
a pioneer of {\it Herschel} and of astrochemistry.}
\tnotetext[Herschel-Acknowledgment]{{\it Herschel} is an ESA space observatory 
with science instruments provided by European-led Principal Investigator consortia 
and with important participation from NASA.}


\author[Koeln]{Holger S.P.~M\"uller\corref{cor}}
\ead{hspm@ph1.uni-koeln.de}
\cortext[cor]{Corresponding author.}
\author[Madrid]{Jos\'e Cernicharo}
\author[Madrid,Meudon]{M. Ag\'undez}
\author[Leuven,Amsterdam]{L.~Decin}
\author[Paris]{P. Encrenaz}
\author[JPL]{J.C. Pearson}
\author[ESAC]{D.~Teyssier}
\author[Leuven,Amsterdam]{L.B.F.M. Waters}

\address[Koeln]{I.~Physikalisches Institut, Universit{\"a}t zu K{\"o}ln, 
   Z{\"u}lpicher Str. 77, 50937 K{\"o}ln, Germany}
\address[Madrid]{Departamento de Astrof\'isica, Centro de Astrobiolog\'ia, CSIC-INTA, 
  Ctra. de Torrej\'on a Ajalvir km 4, Torrej\'on de Ardoz, 28850 Madrid, Spain}
\address[Meudon]{LUTH, Observatoire de Paris-Meudon, 5 Place Jules Janssen, 
  92190 Meudon, France}
\address[Paris]{LERMA and UMR 8112 du CNRS, Observatoire de Paris, 
  61 Av. de l'Observatoire, 75014 Paris, France}
\address[Leuven]{Instituut voor Sterrenkunde, Katholieke Universiteit Leuven, 
  Celestijnenlaan 200D, 3001 Leuven, Belgium}
\address[Amsterdam]{Astronomical Institute Anton Pannekoek, University of Amsterdam, 
  Science Park XH, Amsterdam, The Netherlands}
\address[JPL]{Jet Propulsion Laboratory, 4800 Oak Grove Drive, MC 168-314, 
  Pasadena, CA 91109 U.S.A.}
\address[ESAC]{European Space Astronomy Centre, ESA, P.O. Box 78, 
  28691 Villanueva de la Ca\~nada, Madrid, Spain}

\begin{abstract}
A molecular line survey has been carried out toward the carbon-rich asymptotic 
giant branch star CW~Leo employing the HIFI instrument on board of the 
{\it Herschel} satellite. 
Numerous features from 480~GHz to beyond 1100~GHz could be assigned unambiguously 
to the fairly floppy SiC$_2$ molecule. However, predictions from laboratory data 
exhibited large deviations from the observed frequencies even after some 
lower frequency data from this survey were incorporated into a fit. 
Therefore, we present a combined fit of all available laboratory data together 
with data from radio-astronomical observations. 
\end{abstract}


\end{frontmatter}




\section{Introduction}
\label{introduction}

Silacyclopropynylidene, SiC$_2$, somewhat better known as silicon dicarbide, 
is a fascinating molecule for spectroscopists, structural and quantum chemists 
as well as astronomers. In 1926, uncataloged bands near 500~nm were discovered 
in the spectra of several carbon-rich asymptotic giant branch (AGB) 
stars~\cite{SiC2_A-X_astro_1926a,SiC2_A-X_astro_1926b}. These are late-type stars 
which produce elements heavier than helium and which eject large quantities of 
gaseous material as well as dust which form a circumstellar envelope (CSE). 
Thirty years later, laboratory spectroscopy established that the molecule 
SiC$_2$ is the carrier of these bands~\cite{SiC2_A-X_lab-ident_1956}. 
It was assumed that the molecule has a linear SiCC structure (silapropadienediylidene) 
in analogy with the isoelectronic propadienediylidene, C$_3$. 
Although some later studies cast doubt on the linear structure of the molecule, 
it took almost another 30 years until the analysis of the rotational structure 
of the electronic origin band unmistakably determined the structure as 
silacyclopropynylidene~\cite{SiC2_A-X_lab-non-lin_1984}. In the course 
of their analysis, the authors instigated quantum chemical calculations 
which provided evidence that the cyclic isomer of SiC$_2$ may be lower 
in energy than the linear form~\cite{SiC2_ai_1984}. 

A plethora of quantum chemical calculations on various properties of SiC$_2$ 
have been published later, yielding energy differences between the linear 
and the cyclic form which depended strongly on the level of the calculation 
and the size of the basis set. 
A high level ab initio calculation concluded that the cyclic isomer of 
SiC$_2$ is the only minimum on the potential energy surface and that the linear 
transition state is 24.3~kJ/mol higher in energy~\cite{SiC2_ai_1997}. However, the 
authors attached a caveat to this value as an anharmonic force field calculation 
provided a much too small value for the vibrational energy of $\varv_3 = 1$ and 
much too large anharmonicity constants. Therefore, some of the authors revisited 
the problem of the energy difference between the two SiC$_2$ structures, the last time 
in 2003 when very high level calculations combined with very large basis sets, basis 
set extrapolation to infinite size as well as additional corrections yielded a value 
of 26.5~kJ/mol~\cite{SiC2_ai_2003}.

The SiC$_2$ molecular parameters obtained in Ref.~\cite{SiC2_A-X_lab-non-lin_1984} 
laid the foundation for progress in laboratory spectroscopy. The $J = 1 - 0$ rotational 
transition frequencies of the three isotopologs SiC$_2$\footnote{Unlabeled atoms refer 
to $^{12}$C and $^{28}$Si.}, $^{29}$SiC$_2$, and $^{30}$SiC$_2$, as well as 
the permanent electric dipole moment were measured using Fourier transform 
microwave spectroscopy~\cite{SiC2_1-0_dip_1989}. 
Subsequently, 34 additional transition frequencies were measured for the 
main isotopic species between 93 and 370~GHz~\cite{SiC2_rot_1989}. 
Even though a comparatively large number of 15 spectroscopic parameters of a 
standard Watson-type Hamiltonian in the $A$-reduction, all parameters 
up to sixth order, were employed in the fit, the transition frequencies 
were reproduced on average to only four times the experimental uncertainties. 

Similarly large bodies of laboratory transition frequencies were obtained 
for SiC$^{13}$C between 339 and 405~MHz~\cite{SiCC-13_isos_astro_1991} 
and, only very recently, for $^{29}$SiC$_2$ and $^{30}$SiC$_2$ between 140 
and 360~GHz~\cite{Si-29_30-C2_rot_2011}.

Higher excited vibrational states have been studied for the main isotopolog. 
Rotational transitions in its low-lying $\varv_3 = 1$ vibrational state between 
186 and 399~GHz~\cite{SiC2_001_rot_1991} and in $\varv_3 = 1$ and 2 were obtained 
between 140 and 400~GHz~\cite{SiC2_001_002_rot_1994}. The symmetry of $\varv_3 = 1$ 
is $b_2$ and can be viewed as an asymmetric bending state which facilitates 
internal rotation of the C$_2$ unit with respect to the Si atom. Its 
vibrational energy has been determined as 196.37~cm$^{-1}$ from an investigation 
into the laser-induced and the dispersed fluorescence of a jet-cooled sample of 
SiC$_2$~\cite{SiC2_LiF_DF_1991}.

Several attempts have been made to model rotational and sometimes also 
rovibrational data to a varying extent and accuracy. A semirigid bender (SRB) 
Hamiltonian was employed~\cite{SiC2_modeling_1994} to reproduce $\varv_3 = 0$ 
and 1 rotational transition frequencies~\cite{SiC2_rot_1989,SiC2_001_rot_1991} 
as well as rovibrational data from their previous~\cite{SiC2_LiF_DF_1991} 
and present work~\cite{SiC2_modeling_1994}. The analysis suggested that 
the linear configuration is not a local minimum, and the cyclic form is 22.5~kJ/mol 
lower than the linear form; the estimated uncertainty was 2.4~kJ/mol. 
They found the energy difference to be in good agreement with results from 
their own ab initio calculations; their highest level value being 21.8~kJ/mol. 
The reproduction of the vibrational data was reasonable, that of 
the rotational data was only qualitative and thus of no use for 
radio-astronomical purposes.

The $\varv_3 = 0$ and 1 state rotational data~\cite{SiC2_rot_1989,SiC2_001_rot_1991} 
were also modeled with a dedicated internal rotation Hamiltonian~\cite{SiC2_modeling_1993}. 
A greatly improved reproduction, within 1.5~times the experimental uncertainties, 
was achieved for the ground vibrational state, albeit at the expense of 16 spectroscopic 
parameters compared to 15 previously~\cite{SiC2_rot_1989}. The barrier to linearity was 
derived as $\sim$54~kJ/mol, more than twice the value from ab initio calculations 
of that time. 

A reasonably successful result has been obtained by employing a conventional 
Watson-type Hamiltonian in the $S$-reduction~\cite{IRC_10216_2008}. The reproduction 
of the ground state rotational data~\cite{SiC2_1-0_dip_1989,SiC2_rot_1989} 
was converged after varying 17 spectroscopic and keeping one fixed. 
The data were reproduced on average as well as in the fit which employed an 
internal rotation Hamiltonian~\cite{SiC2_modeling_1993}; the number of varied parameters, 
however, was larger by still another one. After scaling of the parameters, the 
transition frequencies of $^{29}$SiC$_2$ and $^{30}$SiC$_2$ available at that time, 
mostly from astronomical observation~\cite{SiCC-13_isos_astro_1991,IRC_10216_2008}, 
could be reproduced well after releasing only 5 spectroscopic parameters~\cite{IRC_10216_2008}; 
13 parameter were released in the fit for SiC$^{13}$C because of the large amount 
of accurate laboratory data~\cite{SiCC-13_isos_astro_1991}.

The unambiguous assignment of a spectroscopic feature observed in space relies 
on reliable predictions which are usually based on laboratory data which, 
for the most part, have been obtained in approximately the same frequency domain. 
The predictions may also be based in part on data from astronomical observations, 
recent examples include H$^{13}$CO$^+$~\cite{HC-13-O+_2004}, 
DCO$^+$~\cite{DCO+_2005,DCO+_etc_2009}, DNC and HN$^{13}$C~\cite{DCO+_etc_2009}, 
C$_2$H~\cite{C2H_2009}, and C$^{13}$CH~\cite{CC-13-H_2010}. Sometimes, 
identifications are even possible in the absence of laboratory data, as 
demonstrated recently by the detection of C$_5$N$^-$ in the CSE of CW~Leo~\cite{det_C5N-}.
Frequencies have to be predicted with accuracies better than around one tenth 
of the line width to permit extraction of dynamical information. Some type of 
intensity information, such as that at a certain temperature, the line strength, 
or the Einstein $A$-value is needed as is additional auxiliary information such as 
quantum numbers, lower or upper state energies etc. Pickett's {\scriptsize SPCAT} 
and {\scriptsize SPFIT} programs \cite{Herb} have been developed for that purpose 
and have evolved over time, see e.\,g. Ref.~\cite{editorial_Herb-Ed}. 
These programs are routinely used in the Cologne Database for Molecular 
Spectroscopy\footnote{Internet address: https://cdms.astro.uni-koeln.de/classic/}, 
CDMS~\cite{CDMS_1,CDMS_2}, to provide in its catalog section\footnote{Internet address: 
https://cdms.astro.uni-koeln.de/classic/entries/} predictions of (mostly) rotational spectra 
of molecules which may be found in various environments in space.

The spectroscopic parameters from the first rotational analysis of the electronic 
spectrum of SiC$_2$~\cite{SiC2_A-X_lab-non-lin_1984} were accurate enough to 
identify nine unassigned emission features, previously observed between 93 and 171~GHz 
toward the carbon-rich AGB star CW~Leo, also known as IRC~+10216, as belonging to 
SiC$_2$ and improved the SiC$_2$ structural parameters~\cite{SiC2_det-IRC_1984}.
Using the structural parameters from that work, the three $J = 4 - 3$, 
$\Delta K_a = 0$ transitions of $^{29}$SiC$_2$ and $^{30}$SiC$_2$ were detected 
toward the same source very soon thereafter~\cite{Si-29_30-C2_det-IRC_1986}; 
two of these transitions were marginally detected for SiC$^{13}$C. 
Extensive sets of transition frequencies of $^{29}$SiC$_2$, $^{30}$SiC$_2$, and 
SiC$^{13}$C were obtained from radio-astronomical observation between 90 
and 241~GHz together with laboratory rest frequencies for the latter 
isotopolog~\cite{SiCC-13_isos_astro_1991}.

Radio lines of SiC$_2$ have also been detected in the CSEs of other C-rich AGB stars, 
e.\,g. toward II~Lup, which is also known as IRAS 15194-5115~\cite{SiC2_II_Lup_1993}. 
However, SiC$_2$ features are particularly strong toward CW~Leo, which, to a large extent, 
is due to its proximity to our Solar system. CW~Leo has been studied extensively, 
and many molecular species, such as CN$^-$~\cite{det_CN-} and FeCN~\cite{det_FeCN}, 
have been detected toward its CSE exclusively or for the first time, including many 
Si-containing ones, such as SiCN and SiNC~\cite{det_SiNC}. 
The source is a subject of observations in several key projects carried out with 
the recently launched {\it Herschel} satellite~\cite{Herschel}. One of these projects 
is a molecular line survey carried out with the Heterodyne Instrument for the 
Far-Infrared (HIFI)~\cite{HIFI}. This high-resolution instrument covers, 
in several bands, the 480$-$1250 and 1410$-$1910~GHz regions. 
A preliminary analysis of a lower frequency region (554$-$637~GHz) revealed 
that a rather small number of molecules account for a large fraction 
of the emission features~\cite{SiC2_HIFI_2010}. These molecules are CS, SiO, SiS, 
HCN, and, in particular, SiC$_2$. Predictions of the rotational spectrum 
based on Ref.~\cite{IRC_10216_2008} turned out to be very good for transitions 
with higher values of $K_a$, but showed increasing deviations of up to 10~MHz 
for transitions with decreasing $K_a$, contrary to common expectations. 
A combined fit of these transitions frequencies together with those from 
laboratory spectra~\cite{SiC2_1-0_dip_1989,SiC2_rot_1989} required only 
one additional parameter~\cite{SiC2_HIFI_2010} to reproduce the astronomical data 
within uncertainties, and the laboratory data about as well as 
before~\cite{IRC_10216_2008}. Predictions of the SiC$_2$ rotational spectrum based 
on these results are currently available as version~2 in the 
CDMS\footnote{https://cdms.astro.uni-koeln.de/classic/entries/archive/SiC2/28SiC2/001 \_002.vers2/c052528.cat; 
see https://cdms.astro.uni-koeln.de/classic/entries /archive/SiC2/28SiC2/001\_002.vers2/e052528.cat 
for the documentation}. 
Since they were still not appropriate to predict the observed emission features 
satisfactorily to high frequencies (beyond 1100~GHz), we present here a combined fit 
using these as well as laboratory data, supplemented with data from additional 
astronomical observations.


 \begin{figure*}
 \begin{center}
  \includegraphics[width=15.0cm]{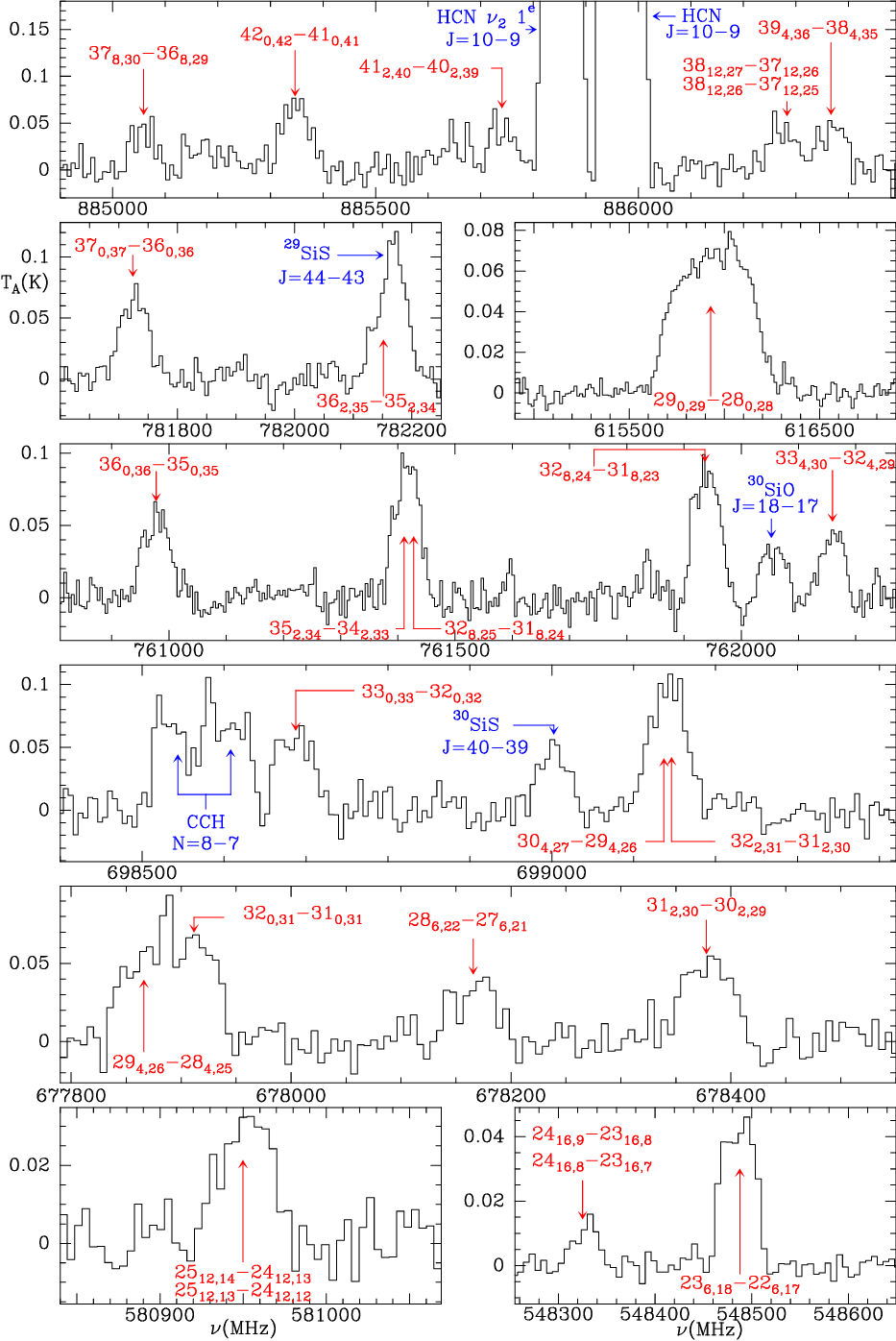}
 \end{center}
  \caption{Sections of the molecular line survey of CW~Leo displaying selected SiC$_2$ 
            transitions. Their quantum numbers $J'_{K'_a,K'_c} - J''_{K''_a,K''_c}$ are 
            given in red (grey). The formulae and quantum numbers of other lines are 
            given in blue (black).}
  \label{spectrum}
 \end{figure*}


\section{Observed spectrum}
\label{obs_spectrum}

Silacyclopropynylidene, SiC$_2$, is a triangular, fairly asymmetric rotor with 
$\kappa = (2B - A - C)/(A - C) = -0.7117$. 
Its dipole moment of 2.393~(6)~D~\cite{SiC2_1-0_dip_1989} is along the $a$-axis. 
Only even $K_a$ exist because of the two equivalent C nuclei with zero spin. 
$R$-branch transitions ($\Delta J = J' - J'' = +1$) with $\Delta K_a = 0$ are 
the strongest ones, and most of the transitions observed in the 
laboratory~\cite{SiC2_1-0_dip_1989,SiC2_rot_1989} and in space, see e.\,g. 
Refs.~\cite{SiC2_det-IRC_1984,Si-29_30-C2_det-IRC_1986,SiCC-13_isos_astro_1991}, 
fall into this category. $Q$-branch transitions with $\Delta K_a = 0$ and transitions 
with $\Delta K_a = \pm2$ also have considerable intensities because of $\kappa$ being 
very different from $-1$. The former type of transitions was detected in space, 
the latter both in the laboratory~\cite{SiC2_rot_1989} and in space.

The line survey of CW~Leo was carried out with the HIFI instrument~\cite{HIFI} on board 
of the {\it Herschel} satellite~\cite{Herschel} May 11$-$15 2010 employing all six 
bands. Some details on the observations and on the data reduction have been given 
earlier~\cite{SiC2_HIFI_2010}, and additional information will be provided in a 
subsequent manuscript on the whole line survey. The lower resolution mode provides 
a point spacing of about 1.1~MHz which is well sufficient even at low frequencies. 
A fair fraction of the moderately strong to weak lines in the four lowest bands 
from 480 to 1120~GHz could be assigned to SiC$_2$. No lines were observed with 
sufficient signal-to-noise ratio in bands 5 to 7 (1120$-$1250~GHz and 1410$-$1910~GHz).
The center frequencies were determined from a fit to the observed line profile 
using the CLASS program of the GILDAS package\footnote{See 
\texttt{http://www.iram.fr/IRAMFR/GILDAS}}. 
The rotational temperature of SiC$_2$ was determined as $\sim$204~K; the Boltzmann 
peak at this temperature is near 500~GHz. 261 distinct features have been identified 
in the HIFI spectra corresponding to 319 rotational transitions with $19 \le J \le 53$ 
and $K_a \le 16$. All transitions were $R$-branch transitions with $\Delta K_a = 0$. 
The accuracies of the frequencies range from 1.5~MHz for relatively strong, isolated 
lines at low frequencies to 20~MHz for lines with low signal-to-noise ratio at 
high frequencies. Selected SiC$_2$ transitions from the HIFI molecular line survey 
are shown in Fig.~\ref{spectrum}.

Additional spectral data of IRC~+10216 in the radio domain (90$-$358~GHz) were 
obtained with the IRAM\footnote{IRAM is an international institute for research in 
millimeter astronomy, co-funded by the Centre National de la Recherche Scientifique 
(France), the Max Planck Gesellschaft (Germany) and the Instituto Geografico Nacional 
(Spain).} 30~m telescope, located at Pico Veleta, Granada (Spain). 
Lines in the $\lambda$ 2~mm wavelength domain have been previously 
reported~\cite{2mm_survey}, while those lying at $\lambda$ 3~mm and 1~mm will be 
presented in detail in forthcoming papers. The center frequencies were determined 
in the same way as used for the lines observed with Herschel. 
A total of 59 SiC$_2$ lines were clearly observed in emission with a spectral resolution 
of 1$-$1.25~MHz below 260~GHz and of 2~MHz above this frequency. 
The 59 lines correspond to 75 rotational transitions with $K_a \le 12$. 
Most of them are $R$-branch transitions with $\Delta K_a = 0$, but there were also 
five $Q$-branch transitions ($\Delta J = 0$) with $K_a = 2 - 0$ 
and two with $K_a = 2 - 2$. 
Although the observed lines are relatively wide, around 29~km\,s$^{-1}$ in equivalent 
radial velocity, their edges are very sharp with a small broadening at the base due 
to micro turbulence velocities of 1.5~km\,s$^{-1}$. Hence, lines with a high 
signal-to-noise ratio can be fitted with an accuracy better than 1~km\,s$^{-1}$. 
Consequently, the experimental accuracies on the frequencies derived towards IRC~+10216 
depend on the line intensity, being around 1~MHz for the weakest lines and around 
0.1~MHz for the strongest ones.

The transition frequencies from the HIFI and IRAM 30~m observations are given in the 
supplementary material together with those from laboratory experiments with quantum numbers, 
uncertainties, and residuals o$-$c between observed frequencies and those calculated 
from the final fit.


\section{Spectroscopic analysis}
\label{analysis}

The choice of spectroscopic parameters is often not unique if one has to reproduce 
a large set of transition frequencies, in particular in the case of floppy molecules, 
such as SiC$_2$. One way of getting a comparatively small and possibly even unique 
data set is to examine carefully which parameter improves the quality of the fit the most, 
as judged by the rms error, also known as reduced $\chi ^2$, and if one includes and 
keeps only parameters determined with a certain level of significance. 
However, this procedure may run into difficulties if parameters of a given order have 
rather similar magnitudes or if certain parameters are strongly correlated. 
Moreover, a particularly small parameter set or one yielding the smallest rms error 
does not always yield the most reliable predictions, in particular if the amount 
of transition frequencies is rather small.

As indicated in section~\ref{introduction}, the SiC$_2$ laboratory data from 
Ref.~\cite{SiC2_rot_1989} were reproduced there to only four times the reported 
uncertainties employing Watson's $A$-reduction of the rotational Hamiltonian with a 
complete set of parameters up to sixth order. As will be shown later, the quality of 
the fit is a result of the particular choice of parameters, not a result of the choice 
of the reduction. In that fit, 
the residuals were distributed unevenly, and the $K$-ordering of the energy levels 
was incorrect starting at $J = 26$. Refs.~\cite{SiC2_modeling_1993,IRC_10216_2008} 
used a dedicated internal rotation Hamiltonian with 16 parameters and Watson's 
$S$-reduction of the rotational Hamiltonian with 17 varied parameters, respectively, 
to reproduce these data within about 1.5~times the reported uncertainties. 
This suggests that the laboratory data~\cite{SiC2_rot_1989} were estimated too 
optimistically. Here, as in previous fits~\cite{IRC_10216_2008,SiC2_HIFI_2010}, 
the uncertainties from that work were multiplied with 1.5 and rounded up.

Starting initially with the parameter set from Ref.~\cite{IRC_10216_2008}, 
the complete list of transition frequencies from HIFI observations, from 
laboratory spectroscopy, and from selected additional astronomical observations 
could be reproduced well with the $S$-reduction of the rotational Hamiltonian. 
However, these fits required around 24 varied parameters compared with 17 and 18, 
respectively, in previous fits~\cite{IRC_10216_2008,SiC2_HIFI_2010}. 
Therefore, attempts were also made to fit the data employing the $A$-reduction. 
In one fit, the procedure described above was followed; a fit with only 18 spectroscopic 
parameters achieved an rms error of 0.77. However, comparatively many off-diagonal 
distortion parameters, such as $\phi_K$, $l_K$, and $p_K$, were used in the fit. 
Moreover, the partition function values, calculated by including energy level up 
to $J = 90$ and $K_a = 40$ for convergence at 300~K, differed remarkably little 
between 10~K and 300~K. This was caused by decreasing energies, eventually turning 
negative, at low $K_a$. Since such energies are unphysical, the fit was 
therefore discarded. 

In a subsequent fit, off-diagonal distortion parameters were avoided as far 
as possible. The fit achieved an overall rms error of 0.73 with 20 parameters. 
Three highly $K$-dependent off-diagonal distortion parameters and $\varPhi_K$, 
used in the previous fit, were replaced by $l_{JK}$ and some additional diagonal 
distortion parameters. These parameters yielded partition function values 
similar to previous works~\cite{IRC_10216_2008,SiC2_HIFI_2010}. 
The laboratory data from Ref.~\cite{SiC2_rot_1989} as well as astronomical observations 
up to 360~GHz were reproduced with partial rms errors marginally smaller than 1.0, 
while the transition frequencies obtained from HIFI observations had a partial rms error 
of 0.63, suggesting that, overall, they may have been judged slightly conservatively. 
The parameter $\varPhi_K$, determined insignificantly in the previous laboratory spectroscopic 
work~\cite{SiC2_rot_1989} and judged to be rather large in a very recent laboratory study 
of $^{29}$SiC$_2$ and $^{30}$SiC$_2$~\cite{Si-29_30-C2_rot_2011}, was still not determined 
with significance in later fits and was therefore not included in the final fit.
The resulting spectroscopic parameters are given in Table~\ref{SiC2-spec-parameters}.
The change in reduction of the rotational Hamiltonian prompted an attempt to fit 
the laboratory data~\cite{SiC2_1-0_dip_1989,SiC2_rot_1989} with an $A$ reduction parameter 
set. All transition frequencies from astronomical observations were eliminated from 
the line list, and all parameters with uncertainties not much smaller than the value 
were eliminated up to before the point at which the rms error increased substantially. 
The final fit consisted of 15 parameters which reproduced the data as well as the fits 
described in Refs.~\cite{SiC2_modeling_1993,IRC_10216_2008}, but actually with one 
and two varied parameters, respectively, less. The spectroscopic parameters 
from that fit have also been given in Table~\ref{SiC2-spec-parameters}.


\begin{table}
  \begin{center}
  \caption{Spectroscopic parameters$^a$ (MHz) for SiC$_2$ obtained in the present 
  investigation.}
  \label{SiC2-spec-parameters}
{\footnotesize
  \begin{tabular}{lr@{}lr@{}l}
  \hline 
Parameter & \multicolumn{2}{l}{All data} & \multicolumn{2}{l}{Lab. data only$^b$} \\
\hline
$A-(B+C)/2$                &  40673&.821\,(37)    &  40674&.109\,(60)    \\
$(B+C)/2$                  &  11800&.14670\,(66)  &  11800&.14722\,(89)  \\
$(B-C)/4$                  &    679&.28139\,(53)  &    679&.28365\,(87)  \\
$\varDelta_K$              &   $-$1&.2841\,(89)   &   $-$1&.2092\,(153)  \\
$\varDelta_{JK}$           &      1&.538195\,(69) &      1&.538100\,(86) \\
$\varDelta_J \times 10^3$  &     13&.1962\,(28)   &     13&.2188\,(38)   \\
$\delta_K \times 10^3$     &    869&.88\,(20)     &    870&.38\,(29)     \\
$\delta_J \times 10^3$     &      2&.41187\,(170) &      2&.42028\,(374) \\
$\varPhi_{KJ} \times 10^6$ &    381&.0\,(33)      &    426&.9\,(102)     \\
$\varPhi_{JK} \times 10^6$ &  $-$48&.14\,(81)     &  $-$61&.19\,(308)    \\
$\varPhi_J \times 10^9$    &  $-$84&.9\,(36)      &       &              \\
$\phi_K \times 10^3$       &      1&.084\,(16)    &      0&.824\,(59)    \\
$\phi_{JK} \times 10^6$    &  $-$33&.51\,(43)     &  $-$29&.96\,(113)    \\
$L_{KKJ} \times 10^9$      &    319&.6\,(225)     &    135&.3\,(207)     \\
$L_{JK} \times 10^9$       & $-$148&.4\,(43)      &  $-$92&.9\,(69)      \\
$L_{JJK} \times 10^9$      &   $-$1&.43\,(31)     &   $-$5&.94\,(72)     \\
$l_{JK} \times 10^9$       &   $-$1&.575\,(153)   &       &              \\
$P_{KKJ} \times 10^9$      &   $-$1&.179\,123)    &       &              \\
$P_{KJ} \times 10^{12}$    &    426&.3\,(271)     &       &              \\
$P_{JK} \times 10^{12}$    &  $-$49&.50\,(227)    &       &              \\
\hline \hline
\end{tabular}\\[2pt]
}
\end{center}
$^a$ {\footnotesize
Watson's $A$-reduction was used in the representation $I^r$.
Numbers in parentheses are 1\,$\sigma$ uncertainties in units of the 
least significant figures.\\}
$^b$ {\footnotesize Refs.~\cite{SiC2_1-0_dip_1989,SiC2_rot_1989}}
\end{table}


\section{Discussion}
\label{Discussion}

The transitions of SiC$_2$, whose frequencies have been measured in the laboratory or 
in space, are almost all $R$-branch transitions, and most of them have $\Delta K_a = 0$. 
Spectroscopic parameters obtained from such a data set permit reasonable extrapolation 
up to around twice the highest frequencies in the line list even under favorable conditions. 
Favorable conditions are a good quantum number coverage (several transition frequencies 
for each spectroscopic parameter with a sufficiently representative range of $J$ and 
$K_a$ quantum numbers) and a Hamiltonian that converges sufficiently fast. 
A predicted transition frequency is deemed to be reasonable in this context 
if the measured line is found within three to five times the predicted uncertainties. 
The SiC$_2$ molecule is comparatively floppy. Moreover, the laboratory data set for 
the main isotopolog of SiC$_2$ is rather sparse and extends to 370~GHz. Nevertheless, 
the highest frequency transition in Ref.~\cite{SiC2_HIFI_2010}, which report a 
preliminary analysis of HIFI data, is $30_{0,30} - 29_{0,29}$ at 636346~MHz, and it 
shows among those lines the largest deviation from the predictions based on 
Ref.~\cite{IRC_10216_2008} with almost 10~MHz. However, this deviation corresponds to 
only 5.5~times the predicted uncertainty. Moreover, the frequency from the final analysis 
has been corrected to a slightly lower frequency, in better agreement with 
the initial predictions. 

Larger deviations from the predictions were observed around 1~THz, but these had to be 
expected since the predicted transition frequencies have uncertainties of several tens 
to almost 100~MHz. In fact, predictions from Refs.~\cite{IRC_10216_2008,SiC2_HIFI_2010} 
are quite reasonable for the most part even if the deviations are large enough 
to be easily recognizable in the astronomical spectra.

Fits employing Watson's $A$-reduction of the rotational Hamiltonian reproduce the 
experimental transition frequencies with fewer parameters than those using the 
$S$-reduction and should thus be preferred as long as the energy ordering is correct 
up to sufficiently high quantum numbers. 
It may be interesting to compare the quality of the predictions of the transitions 
detected with HIFI from either laboratory fit from Refs.~\cite{IRC_10216_2008} 
or in Table~\ref{SiC2-spec-parameters}. Predictions based on $S$-reduction 
parameters~\cite{IRC_10216_2008} are quite good for transitions having higher values 
of $K_a$ ($6 - 14$), but deteriorate for lower $K_a$ values. In contrast, predictions 
based on $A$-reduction parameters (Table~\ref{SiC2-spec-parameters} under heading 
''Lab. data only'') are quite good for low-$K_a$ transition, but deteriorate 
for higher $K_a$ values, which is closer to common expectations. 
Both fits as well as both predictions, as well as additional fits are available 
in the archive section of the CDMS\footnote{Internet address: 
https://cdms.astro.uni-koeln.de/classic/entries/archive/SiC2/}. 
Overall, however, the quality of the predictions is quite similar and do not provide 
a reasoning which reduction should be preferred. A very recent laboratory study of 
$^{29}$SiC$_2$ and $^{30}$SiC$_2$~\cite{Si-29_30-C2_rot_2011} also obtained fits 
of similar quality for the $A$- and the $S$-reduction, but only if one more parameter 
was used in the $S$-reduction and preferred thus the $A$-reduction. 

The SiC$_2$ spectroscopic parameters in Table~\ref{SiC2-spec-parameters} from laboratory 
data only and from all data agree quite well as far as they have been determined, 
in particular for the lower order parameters, rotational and quartic distortion constants. 
The higher order parameters show larger deviations, possibly reflecting the floppy nature 
of the molecule. However, the deviations are in all instances, at most, barely significant, 
meaning larger than three times the initial uncertainties.

The measured transition frequencies have been reproduced well with parameter sets of 
conventional Watson-type Hamiltonians. The molecule is rather floppy already in its 
ground vibrational state such that a fairly large number of parameters were used. 
Additionally, the magnitudes of the parameters decrease slowly such that the contributions 
of the higher order parameters are fairly large at energy levels having high values 
of $J$ or $K_a$. Extrapolations to slightly higher quantum numbers may be reasonable, 
but should be viewed with caution.

The dedicated internal rotation Hamiltonian~\cite{SiC2_modeling_1993} did not achieve 
a considerably more compact representation of the measured transition frequencies 
known at that time. Without any available predictions for transition frequencies or 
rotational energies it seems questionable if that model is more appropriate than a 
conventional Watson-type Hamiltonian, even more so as the predicted barrier to linearity 
was rather far from values from ab initio calculations~\cite{SiC2_ai_2003} or 
from the SRB extrapolation from highly vibrationally excited levels~\cite{SiC2_modeling_1994}. 
Moreover, one may question if an internal rotation model is appropriate at all. 
On the other hand, the semirigid bender approach~\cite{SiC2_modeling_1994} may be 
a viable alternative, but only if it will achieve reproduction of the 
measured transition frequencies to within estimated uncertainties with a parameter set 
that is not much larger than those used previously.

Additional laboratory data for the main isotopolog of SiC$_2$ will be of great use in 
particular if these connect levels with $K_a \geq 2$ via transitions with 
$\Delta K_a \geq 2$. But even $Q$-branch transitions with $\Delta K_a = 0$ may be 
very useful as they sample higher $J$ levels at comparatively low frequencies. 
Such higher excited transitions may well be observable with existing interferometric 
facilities such as PdBI, ATCA, or SMA, and even more so with upcoming instruments 
such as ALMA, NOEMA, or possibly EVLA.


\section{Conclusions}
\label{Conclusions}

Astronomical observations with {\it Herschel} have been used to improve spectroscopic 
parameters of SiC$_2$ in its ground vibrational state greatly. The transition frequencies 
obtained from laboratory data as well as from astronomical observations have been reproduced 
to within the estimated uncertainties employing a conventional Watson-type Hamiltonian. 
These data should be useful to test alternative models to describe the rotational or 
even rovibrational energy levels of this floppy molecule. Predictions of the rotational 
spectrum as well as line, parameter and other auxiliary files, both from present fits 
as well as previous ones, will be available in the CDMS~\cite{CDMS_1,CDMS_2}.



\section*{Acknowledgements}

H.S.P.M. is very grateful to the Bundesministerium f\"ur Bildung und 
Forschung (BMBF) for financial support aimed at maintaining the 
Cologne Database for Molecular Spectroscopy, CDMS. This support has been 
administered by the Deutsches Zentrum f\"ur Luft- und Raumfahrt (DLR). 
J.C. thanks the Spanish MICINN for funding support under grants AYA2009-07304 
and CSD-2009-00038.

\appendix

\section*{Appendix A. Supplementary Material}

Supplementary data associated with this article can be found, in
the online version, at doi: 10.1016/j.jms.2011.11.006.



\bibliographystyle{elsarticle-num}
\bibliography{<your-bib-database>}




\end{document}